\documentclass{aip-cp}

\usepackage[numbers]{natbib}
\usepackage{rotating}
\usepackage{graphicx,epstopdf}
\usepackage{lineno}

\begin{document}

\newcommand{\legend}       {LEGEND}
\newcommand{\gerda}       {GERDA}
\newcommand{\mjd}       {\textsc{Majorana Demonstrator}}

\def\nuc#1#2{${}^{#1}$#2}
\def\mee{$\langle m_{\beta\beta} \rangle$}
\def\mnu{$\langle m_{\nu} \rangle$}
\def\ml{$m_{lightest}$}
\def\gnu{$\langle g_{\nu,\chi}\rangle$}
\def\mmod{$\| \langle m_{\beta\beta} \rangle \|$}
\def\mb{$\langle m_{\beta} \rangle$}
\def\BBz{$\beta\beta(0\nu)$}
\def\zBB{$0\nu\beta\beta$}
\def\BBm{$\beta\beta(0\nu,\chi)$}
\def\BBt{$\beta\beta(2\nu)$}
\def\nonubb{$\beta\beta(0\nu)$}
\def\twonubb{$\beta\beta(2\nu)$}
\def\BB{$\beta\beta$}
\def\Mz{$M_{0\nu}$}
\def\Mt{$M_{2\nu}$}
\def\MzG{$M^{GT}_{0\nu}$}           
\def\MzF{$M^{F}_{0\nu}$}                
\def\MtG{$M^{GT}_{2\nu}$}           
\def\MtF{$M^{F}_{2\nu}$}                
\def\Tz{$T^{0\nu}_{1/2}$}
\def\Tt{$T^{2\nu}_{1/2}$}
\def\Tc{$T^{0\nu\,\chi}_{1/2}$}
\def\Rz{$\Gamma_{0\nu}$}            
\def\Rt{$\Gamma_{2\nu}$}            
\def\ms{$\delta m_{\rm sol}^{2}$}
\def\ma{$\delta m_{\rm atm}^{2}$}
\def\mot{$\delta m_{12}^{2}$}
\def\mtt{$\delta m_{23}^{2}$}
\def\ts{$\theta_{\rm sol}$}
\def\ta{$\theta_{\rm atm}$}
\def\tot{$\theta_{13}$}
\def\gpp{$g_{pp}$}                  
\def\qval{$Q_{\beta\beta}$}                 
\def\Gerda{{\sc Gerda}}             
\def\MJ{{\sc Majorana}}             
\def\DEM{{\sc Demonstrator}}             
\def\be{\begin{equation}}
\def\ee{\end{equation}}
\def\cpRty{counts/(ROI t y)}
\def\onecpRty{1~counts/(ROI t y)}
\def\fourcpRty{4~counts/(ROI t y)}
\def\threecpRty{3~counts/(ROI t y)}
\def\ppc{P-PC}                          
\def\nsc{N-SC}                          
\def\cosixty{$^{60}\mathrm{Co}$}
\def\gess{$^{76}\mathrm{Ge}$}
\def\tltze{$^{208}\mathrm{Tl}$}
\def\thttt{$^{232}\mathrm{Th}$}
\def\utte{$^{238}\mathrm{U}$}
\def\mubqkg{$\mu\mathrm{Bq/kg}$}
\def\cusulfate{$\mathrm{CuSO}_4$}
\def\LSGe{LSGe}

%
\title{The Large Enriched Germanium Experiment for Neutrinoless Double Beta Decay (LEGEND)}

%
\author[LBNL]{N.~Abgrall}
\author[ITEP]{A.~Abramov}
\author[IKZ]{N.~Abrosimov}
\author[MPIM]{I.~Abt}
\author[GSSI]{M.~Agostini}
\author[Dokuz]{M.~Agartioglu}
\author[Dokuz]{A.~Ajjaq}
\author[UW]{S.I.~Alvis}
\author[USC,ORNL]{F.T.~Avignone III}
\author[SDSMT]{X.~Bai}
% ...

\affil[LBNL]{Institute for Nuclear and Particle Astrophysics and Nuclear Science Division, Lawrence Berkeley National Laboratory, Berkeley, California 94720}
\affil[ITEP]{National Research Centre ``Kurchatov Institute'', Moscow}
\affil[IKZ]{Leibniz Institute for Crystal Growth, Berlin}
\affil[MPIM]{Max-Planck-Institut f\"{u}r Physik, M\"{u}nchen}
\affil[GSSI]{Gran Sasso Science Institute, L'Aquila}
\affil[Dokuz]{Department of Physics, Dokuz Eyl\"{u}l University, Buca, Izmir}
\affil[UW]{Center for Experimental Nuclear Physics and Astrophysics, and Department of Physics, University of Washington, Seattle, Washington 98195}
\affil[USC]{Department of Physics and Astronomy, University of South Carolina, Columbia, South Carolina 29208}
\affil[ORNL]{Oak Ridge National Laboratory, Oak Ridge, Tennessee 37830}
\affil[SDSMT]{South Dakota School of Mines and Technology, Rapid City, South Dakota, 57701}
% ...

\corresp[cor1]{Corresponding author: jfw@unc.edu}

\maketitle



\begin{abstract}
The observation of neutrinoless double-beta decay (\zBB) would show that lepton number is violated, reveal that neutrinos are Majorana particles, and provide information on neutrino mass. A discovery-capable experiment covering the inverted ordering region, with effective Majorana neutrino masses of $15 - 50$ meV,  will require a tonne-scale experiment with excellent energy resolution and extremely low backgrounds, at the level of $\sim$0.1~count /(FWHM$\cdot$t$\cdot$yr) in the region of the signal.  The current generation $^{76}$Ge experiments GERDA and the \mjd, utilizing high purity Germanium detectors with an intrinsic energy resolution of 0.12\%, have achieved the lowest backgrounds by over an order of magnitude in the \zBB\ signal region of all \zBB\ experiments.  Building on this success, the LEGEND collaboration has been formed to pursue a tonne-scale $^{76}$Ge experiment. The collaboration aims to develop a phased \zBB\ experimental program with discovery potential at a half-life approaching or at 10$^{28}$ years, using existing resources as appropriate to expedite physics results.
\end{abstract}

\maketitle

\section{INTRODUCTION}

Neutrinos are the only known 
fundamental fermions without electric charge. 
As a consequence they can acquire mass not
only through the standard coupling to the Higgs particle 
but also by additional lepton
number violating operators \cite{mohapatra06}.  
As further consequences, neutrinos will - in general - be their own
anti-particles (Majorana particle) and neutrinoless
double beta ($0\nu\beta\beta$) decay may
exist \cite{mohapatra07,avi08,rodejohann15}: a nucleus (A,Z)
decays to  (A,Z+2) + 2e$^-$. The sum of the
electron energies is here equal to the $Q$ value of the
decay ($Q_{\beta\beta}$). This
is the prime signature for $0\nu\beta\beta$ decay.

\section{OPERATING $^{76}$Ge EXPERIMENTS}

Currently the \gerda\ and \mjd\ experiments are searching 
for $0\nu\beta\beta$ decay
of $^{76}$Ge. Both experiments use germanium detectors made out of 
material with the $^{76}$Ge fraction enriched to at least 87\%.
A major advancement over previous experiments has been GERDA's and \MJ's independent development
and use of p-type point-contact (PPC) high purity germanium (HPGe) detectors~\cite{Luke1989}.
PPC detectors, with signal time evolution similar to drift detectors, have a number of advantages over conventional coaxial HPGe detectors:  simple fabrication and readout,
excellent pulse shape discrimination (PSD) between \zBB\ events and backgrounds, and very low capacitance,
providing a low-energy threshold allowing the reduction of potential background from cosmogenic $^{68}$Ge.
A primary difference between the two experiments is the shielding
used against external radiation.
However, there are also a number of commonalities including
careful attention to backgrounds and the development of ultra-clean fabrication techniques.


The second phase of \gerda\ started in December 2015 with 37 enriched detectors arranged in 6 strings
(total mass  35.6~kg enriched to 87\% $^{76}$Ge) in a 64~m$^3$ liquid argon cryostat, with the argon serving as an internal active veto surrounding the detectors.
The current background index achieved for the GERDA PPC modified (or thick window) Broad Energy Germanium (BEGe) detectors is
$0.7_{-0.5}^{+1.1}\cdot 10^{-3}$ cts/(keV$\cdot$kg$\cdot$yr)
with an energy resolution (FWHM) of 3~keV at $Q_{\beta\beta}$ of 2039 keV and a total efficiency of 0.6.
This corresponds to a projected background at $Q_{\beta\beta}$ of 2.1$^{+3.3}_{-1.5}$ cts/(FWHM$\cdot$t$\cdot$yr).
Based on a total exposure of 34.4 kg$\cdot$yr from Phases I and II,
GERDA sets a lower-limit on the half-life of $5.3\cdot10^{25}$ yr at the 90\% confidence level (sensitivity of $4.0\cdot10^{25}$ yr)~\cite{gerda:2017:nature}.
The \mjd\ array contains  35 (29.7 kg) PPC detectors enriched to 88\% $^{76}$Ge
enclosed in a compact graded shield with inner layers of ultra-clean electroformed copper and an external active muon veto.
The experiment is operating at the Sanford Underground Research Facility (SURF).
The full array has been operating since August 2016.
An analysis of initial \mjd\ data with an exposure of 1.39 kg$\cdot$yr finds a background index of
$1.8_{-1.1}^{+3.1}\cdot 10^{-3}$ cts/(keV$\cdot$kg$\cdot$yr)
 with an energy resolution (FWHM) of 2.4~keV at $Q_{\beta\beta}$ and a total efficiency of 0.6.
 This corresponds to a projected background at $Q_{\beta\beta}$ of 4.3$^{+7.5}_{-2.7}$ cts/(FWHM$\cdot$t$\cdot$yr)~\cite{Guiseppe2017}.
 Both experiments project ultimate sensitivities of discovering a signal with a half life of $10^{26}$ years.

\section{NEXT GENERATION REQUIREMENTS and LEGEND}

For $  m_{\beta\beta} = 17$~meV, a typical lower
bound for the inverted neutrino 
mass ordering (90\% range using parameters from \cite{pdg:2016}),
the  `worst-case' half-life for
the most recent $^{76}$Ge shell model calculation is about
$12\cdot 10^{27}$~yr.
The expected number of decays per t$\cdot$yr exposure
will be 0.5.
The current $^{76}$Ge experiments have achieved the best energy
resolution and correspondingly the lowest backgrounds within
an energy window of resolution-FWHM centered at $Q_{\beta\beta}$.
The superior resolution coupled with modest improvements in backgrounds
makes $^{76}$Ge capable of identifying a signal of even a few events at $Q_{\beta\beta}$
and a leading contender to advance to the next generation of tonne-scale $0\nu\beta\beta$ experiments.

\begin{figure}
\centering
\begin{tabular}[b]{c}
\includegraphics[width=0.45\columnwidth]{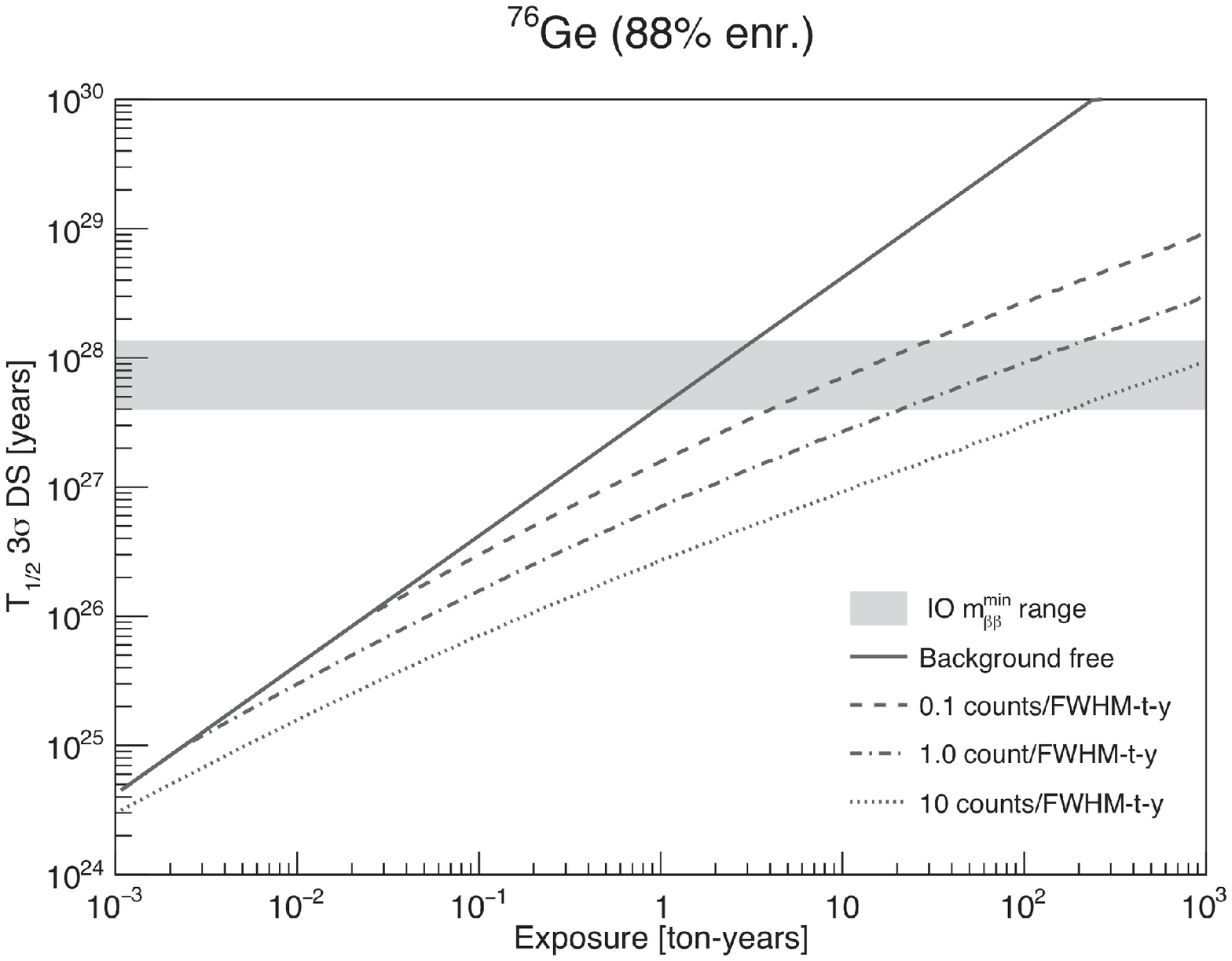} \\
 \small (a)
\end{tabular}
\begin{tabular}[b]{c}
\includegraphics[width=0.45\columnwidth]{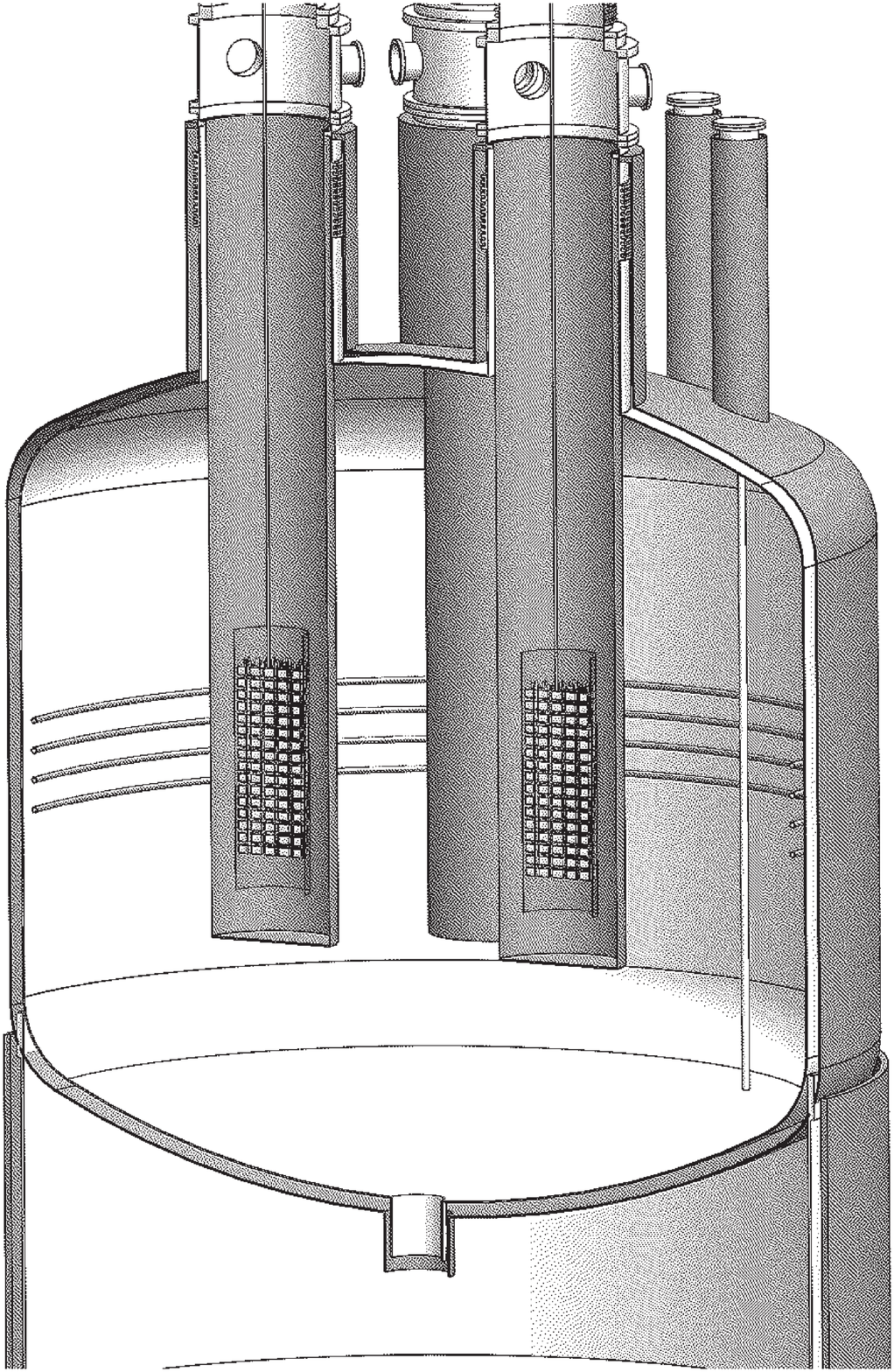} \\
\small (b)
\end{tabular}
\caption{\label{fig:joint} (a) Sensitivity for a signal discovery.  (b) Sketch of baseline cryostat design for \legend-1000.}
\end{figure}

The \legend\ collaboration aims to increase the
sensitivities for $^{76}$Ge in a first phase to $10^{27}$~yr and in a
second phase up to $10^{28}$~yr both for setting a 90\% C.L.~half-life
limit as well as for  ``discovery'' of $0\nu\beta\beta$ decay defined
as a 50\% chance for a signal at 3$\sigma$  significance.
Fig.~\ref{fig:joint}(a) shows the sensitivities of a germanium
experiment for discovery as a function
of the exposure for different background levels. 
A signal efficiency of 0.6 is taken into account.
If the background is ``zero'', sensitivity scales linearly with
the exposure, otherwise only with the square root. For
signal detection a low background is most important since
the transition from linear to square root dependence occurs at
lower exposures, i.e.~for a smaller mean background
count.
Hence the goal is to perform a ``background-free'' measurement, defined as being $<$ 1 mean expected background count at an experimentÕs design exposure.
In \legend's first phase up to 200~kg of Ge detectors, \legend -200, will be operated
in the existing infrastructure of the \gerda\ experiment at the 
Laboratori Nazionali del Gran Sasso (LNGS) in Italy. 
For an exposure of 1~t$\cdot$yr
and under conditions approaching background-free measurement (background index
of 0.6~cts/(FWHM$\cdot$t$\cdot$yr)), the envisioned
half-life sensitivity can be reached. In the following phases a new facility, \legend -1000,
holding up to 1000~kg of 
detectors would be operated with
even lower background of less than 0.1~cts/(FWHM$\cdot$t$\cdot$yr)
and a design exposure above 10 t$\cdot$yr.

\legend\ builds on the success of the current $^{76}$Ge experiments.
Both experiments have achieved
the lowest background level among current 
$0\nu\beta\beta$ experiments
and remain background-free. 
The natural next step towards advancing the sensitivity 
is therefore through the increase of the detector mass
while reducing backgrounds from current levels by a
factor of $\sim$30. 
This further background reduction is necessary
to remain essentially background-free as the total exposure increases.
The combined strength from 
the \gerda\ and \mjd\ concepts and experiences
defines the path towards such envisioned background improvement. 
\legend\ has adopted the \gerda\ design of a
low-Z shielding (water and argon) and an active
veto through the detection of argon scintillation light.
Muon and $\gamma$ induced backgrounds are reduced or vetoed. 
The \mjd\ has achieved a comparable low background level as \gerda\ despite
not having an internal active veto in the detector region. 
This is a consequence of careful selection and control of the radiopurity 
of materials in the vicinity of the target.
In addition, the readout electronics and associated cables yield 
better resolution for the
energy and an improved pulse shape parameter used for background rejection.
The experience and knowledge from \gerda\ and \mjd\, as well as from other \legend\ collaborators with expertise in low-background measurements will be crucial in realizing further background suppression in \legend.

\legend\ aims for a number of
improvements including larger PPC type Ge detectors  with
higher mass via an  ``inverted-coaxial''  \cite{invertedcoax} design,  better readout electronics and higher argon scintillation light
detection efficiency.   
Improvements implemented in \legend -200 should be applicable
for the \legend -1000 or will guide its development.
The superior energy resolution demonstrated by PPC detectors
coupled with improvements in the backgrounds
achieved by GERDA and \mjd\ should position \legend -200
and \legend-1000 as leading experiments in the field.

\section{\legend -200}

\legend -200 plans to operate up to 200~kg of germanium detectors
using the existing \gerda\ infrastructure at LNGS. In order
to be ``background-free'' for an exposure of 1~t$\cdot$yr 
a factor of 5 reduction is needed relative to the latest 
 \gerda\ and \mjd\ background levels.
The existing infrastructure is large enough to house
200~kg of detectors: the neck of the cryostat
has a diameter of 800~mm which is wide enough for
19 strings of detectors with a total outer diameter of 
500-550~mm.
\legend-200 will use the existing \mjd\ and \gerda\ PPC detectors as well as additional new detectors.
The new ones will be of the  inverted-coaxial type
which offer a similar pulse shape performance but can have much higher
mass. Thus the number of channels per kg and the resulting backgrounds from
cables and holders will be reduced compared to the current experiments.

The backgrounds for \gerda\ and \mjd\ are still under evaluation,
but based on a GERDA analysis before cuts, the events near $Q_{\beta\beta}$ are
coming in about equal parts 
from $^{42}$K decays, from degraded $\alpha$ events and
from $^{214}$Bi/$^{208}$Tl decays.  
Pulse shape analysis safely removes
all $\alpha$ decays and we
expect that this will hold true for \legend -200.
The needed background reduction should be achieved by the following measures:
\begin{itemize}
\item  The amount of radio-impurities will be reduced.
This can be realized by using low-mass \mjd\ style components.
\item  An improved design for the scintillation light readout has
proven to detect twice as much light and to result in an improvement
by a similar factor for  $^{214}$Bi rejection.
\item The electronic noise can be reduced such that the
  pulse shape discrimination will be more effective.
\item The mass per detector will increase by a factor of two or more.
 Consequently, the number of cables and holder materials per kg will be
  reduced.
\item For $^{42}$K the $\beta$ particle has to pass through an outer dead layer
   of the germanium where it looses energy such that the fraction
   of events depositing 2039~keV in the active volume is reduced.
   An optimized dead layer thickness will further reduce this background.
   One can also limit the LAr volume contributing to the beta background by encapsulating the detectors in nylon vessels or scintillating plastic material.
\end{itemize}

\section{\legend -1000}

For the next phase of the experiment, \legend-1000,
the exposure of 10~t$\cdot$yr is reached by operating 1000~kg
of detectors for 10 years. This requires new infrastructure and a more ambitious
background goal to remain in the background-free regime.
Several options are still under consideration for \legend-1000, but
an initial baseline design has been established
with bare germanium detectors operating in liquid argon.
Because the enrichment and detector
production will be spread over several years, it is planned
to install the detectors in several batches of $\sim$250~kg
each. The data taking of the already installed detectors 
should continue largely undisturbed. These considerations
lead to the preliminary design shown in Fig.~\ref{fig:joint}(b).
The main cryostat volume is separated by 
thin copper walls from four smaller
volumes of about 3~m$^3$ each.  
Each volume will house a subset of the detectors
and is closed on top by a shutter. 
There will be a lock above each of the shutters
such that the germanium detector array can be
assembled in nitrogen atmosphere in a glove box together
with the argon veto.
An important design criterion will be the minimization
of 'dead' material, i.e.~material like copper or PTFE
which contributes to the background and does not
scintillate. One alternative design being considered is
to use a scintillating plastic, such as polyethylene naphthalate (PEN) as a construction material since it has
good mechanical properties\cite{Majorovits2017}.

Compared to \legend -200  the background needs to be reduced
by another factor of 6. Intrinsic contaminations of the
U/Th decay chains have never been observed in a HPGe 
detector \cite{gerda:2017:bulk}.
External backgrounds from neutrons and $\gamma$s can be 
shielded by liquid argon and water. The most worrisome backgrounds
are summarized below:

\begin{itemize}
\item{\em U/Th contamination in close components:}
 The aim is to increase mass per detector by about a factor of two,
 which will also reduce the fractional number of cables and supports.
 The \mjd\ levels in U/Th for cables and detector support are
 sufficiently low and together with the LAr veto
 these backgrounds should be acceptable.
\item{\em $^{42}$K from $^{42}$Ar:}
 $^{42}$Ar is produced in the air by cosmogenic activation similar
to the production of $^{39}$Ar \cite{ar42production}. 
\legend\ is considering the use of
argon from underground sources based on its potential reduction of backgrounds
and possible availability.  DarkSide has found
that the $^{39}$Ar concentration from an underground source is reduced by
a factor of $1400\pm200$ \cite{darkside:ar39} with comparable reductions expected for $^{42}$Ar.
\item {\em Detector surface contaminations:}
 During detector fabrication surface contaminations like $^{210}$Po
($T_{1/2}=$138~d), $^{210}$Pb ($T_{1/2}=$22.3~yr) or $^{226}$Ra ($T_{1/2}=$1602~yr)
can be introduced.  These backgrounds are effectively identified by PSD for the current measurements.
However a R\&D program is planned to further
reduce the surface contamination for example by reducing the time
detectors are handled in air. 
\item{\em Muon induced background:}
 While the prompt muon background is easily
removed, isotopes like $^{77m}$Ge produced by spallation neutrons
are a potential background source. Time correlation between
muon, neutron capture and background event can reduce this
background. The cosmic-ray flux can be reduced
by a factor of 100 relative 
to that of \legend -200 at LNGS (1400~m rock overburden)
at deep underground laboratories like 
SNOLAB (2000~m rock overburden) in Canada
and CJPL (2400~m rock overburden) in China.
Based on current analysis of results from \gerda\ and \mjd, LNGS and SURF remain potential sites for \legend -1000.
\end{itemize}

\section{Summary}

The \mjd\ and \gerda\ experiments
searching for neutrinoless double beta decay of $^{76}$Ge
using PPC HPGe detectors with intrinsic energy resolution of 0.12\%
have the lowest background levels in terms of FWHM$\cdot$t$\cdot$yr in the field.
These are important prerequisites
for a future signal discovery.
The recently formed \legend\ collaboration will build on
these successes and proposes to probe half-lives approaching
$10^{28}$~yr using 1000 kg of enriched germanium detectors.
In an early phase, 200 kg will be operated in the existing
infrastructure at LNGS with a sensitivity of $10^{27}$yr.
Under favorable funding scenarios \legend -200 could start measurements by 2021.


\bibliographystyle{aipnum-cp}%
\bibliography{legend_medex}%

\end{document}